\documentclass{aa}  
\usepackage{graphicx}
\usepackage{txfonts}
\begin{document}
   \title{The photophoretic sweeping of dust in transient protoplanetary disks}

%   \subtitle{}

   \author{O. Krauss
          \inst{1}
          \and
          G. Wurm
		  \inst{1}
		  \and 
		  O. Mousis
         \inst{2}
           \and 
		  J.-M. Petit
         \inst{2}
           \and 
		  J. Horner
         \inst{3}\fnmsep\thanks{Current address: Astronomy Group, Open University, Walton Hall, Milton Keynes MK7 6AA, UK}
           \and 
		  Y. Alibert
         \inst{3}
          }

   \offprints{O. Krauss}

   \institute{Institut f\"ur Planetologie, University of M\"unster, 
               Wilhelm-Klemm-Str. 10, D-48149 M\"unster, Germany\\
              \email{okrauss@uni-muenster.de}
         \and
             Observatoire de Besan\c{c}on, CNRS-UMR 6091, BP 1615, 25010 Besan\c{c}on Cedex, France
                      \and
             Physikalisches Institut, University of Bern, Sidlerstrasse 5, CH-3012 Bern, Switzerland}

   \date{Received; accepted}

% \abstract{}{}{}{}{} 
% 5 {} token are mandatory
 
\abstract
{Protoplanetary disks start their lives with a dust free inner region
where the temperatures are higher than the sublimation temperature of
solids. As the star illuminates the innermost particles, which are
immersed in gas at the sublimation edge, these particles are subject
to a photophoretic force.}
{We examine the motion of dust particles at the inner edge of
protoplanetary disks due to photophoretic drag.}
{We give a detailed treatment of the photophoretic force for particles
in protoplanetary disks. The force is applied to particles at the
inner edge of a protoplanetary disk and the dynamical behavior of the
particles is analyzed.}
{We find that, in a laminar disk, photophoretic drag increases
the size of the inner hole after accretion onto the central body has
become subdued. This region within the hole becomes an optically
transparent zone containing gas and large dusty particles ($\gg$10
cm), but devoid of, or strongly depleted in, smaller dust
aggregates. Photophoresis can clear the inner disk of dust out to 10
AU in less than 1 Myr. The details of this clearance depend on the
size distribution of the dust. Any replenishment of the dust within
the cleared region will be continuously and rapidly swept out to the
edge. At late times, the edge reaches a stable equilibrium between
inward drift and photophoretic outward drift, at a distance of some
tens of AU. Eventually, the edge will move inwards again as the disk
disperses, shifting the equilibrium position back from about 40 AU to
below 30 AU in 1-2 Myr in the disk model. In a turbulent disk,
diffusion can delay the clearing of a disk by
photophoresis. Smaller and/or age-independent holes of radii of
a few AU are also possible outcomes of turbulent diffusion
counteracting photophoresis.}
{This outward and then inward moving edge marks a region of high dust
concentration. This density enhancement, and the efficient transport
of particles from close to the star to large distances away, can
explain features of comets such as high measured ratios of crystalline
to amorphous silicates, and has a large number of other applications.}


\maketitle
%
%________________________________________________________________

\section{Introduction}

There is virtually no doubt that planetary systems form within
protoplanetary disks, which result as the dense cores of molecular
clouds collapse and new stars accrete in their centre (Bouvier et
al. \cite{bouvier}). Protoplanetary disks initially consist mostly of
gas, with dust making up only $\sim$1\% of the mass. The dust and gas
are often considered to be coupled to each other by this typical 1\%
ratio. This is justified for the interstellar medium, where dust is
only subject to minor size alterations. It is, however, a treacherous
assumption for protoplanetary and circumstellar disks, where particles
are known to rapidly evolve, leading to the formation of large
structures (planetesimals) (Dominik et al. \cite{dominik};
Weidenschilling \cite{weidenschilling2000}). Therefore, it is
important to note at the very beginning of this paper that, within the
scope of this work, the gas and the solids are considered to be two
different components of the protoplanetary disk. In fact, we will show
that they might become partially separated by the effect of
photophoresis for significant time spans over the lifetime of a
disk. According to current statistics detailing the detection of
extrasolar planets and disks, many protoplanetary disks leave behind
planetary systems or debris disks (Greaves et
al. \cite{greaves2006}). What remains nebulous is the timing of
planet, asteroid and comet formation within the lifetime of the
protoplanetary disks (generally a few Myr). If the formation or
shaping processes take place rather late, it is important to know how
the transition from an optically thick gaseous and dusty disk to a
debris disk in which most gas and dust is gone occurs. In two recent
papers, Krauss \& Wurm (\cite{krauss2005}) and Wurm \& Krauss
(\cite{wurm2006}) assumed that the dust might evolve on shorter
timescales than the gas. They considered a late disk, still rich in
gas, but optically thin to radiation from the star. In this case,
photophoresis transports any remaining or replenished dust efficiently
outwards and concentrates it in belts at several tens of AU. Petit et
al. (\cite{petit2006}) showed that in an evolving disk, the
concentration radius moves inwards again. The concentration thus
sweeps back and forth over the early outer Solar system (sweeping
through regions such as the Kuiper belt), and changes the amount of
crystalline silicates which will be added to the bodies in these
regions from a source in the inner Solar system.

The existence of gaseous yet transparent disks is supported by
observations of a class of transitional protoplanetary disks with
large inner holes, such as TW Hya, CoKu Tau/4, DM Tau, and LkCa 15
(Calvet et al. \cite{calvet2002}; Forrest et al. \cite{forrest2004};
D'Alessio et al. \cite{dalessio2005}; Bergin et al. \cite{bergin2004};
Calvet et al. \cite{calvet2005}). Such objects make up about 10\% of
protoplanetary disks (Sicilia-Aguilar et
al. \cite{sicilia2006}). Models imply that they have large inner holes
of several AU radii (Calvet et al. \cite{calvet2002}). These inner
holes are essentially free of dust, with a sharp transition to an
optically thick dusty outer disk. It certainly has to be assumed that
gas is present at the inner edge of the disk, and further outwards
where dust is present. However, even within the region cleared of
dust, gas is obviously present as some of the objects show signs of
active accretion (Najita et al. \cite{najita2006}; Rettig et
al. \cite{rettig2004}). The location of the sharp inner edge as a
function of time can be understood by the application of photophoretic
drift, as we will show in this paper. Photophoresis naturally explains
the clearing of small particles from the inner few AU on rather short
timescales, as soon as an inner hole of radius $\sim$0.5AU (or less)
is present. Photophoresis always acts on particles at the inner edge
of protoplanetary disks. Initially, the radius of the inner edge from
the star might be equal to the sublimation radius. As soon as
accretion decreases sufficiently, photophoresis will (given certain
assumptions) trigger the clearing of protoplanetary disks. The dust
edge will move outwards, away from the sublimation radius. The inner
edge represents a strong enhancement in solid density, and contains
material which is transported over tens of AU from the inner system to
the outer regions. The transportation of material in this way has a
large number of applications. To warrant readability, and not overwork
the manuscript, we merely highlight the applications here and will
publish details separately. For clarity, this paper will remain
focused on the model of the migration of the dusty edge outwards and
inwards again. We present a unified model of small particle migration
based on the ideas of photophoretic drift and concentration quantified
in Krauss \& Wurm (\cite{krauss2005}) and Petit et
al. (\cite{petit2006}). As our example protoplanetary disk, we use an
$\alpha$-disk model (Shakura \& Sunyaev \cite{shakura1973}), taking
into account the effect of photoevaporation (Papaloizou \& Terquem
\cite{papaloizou1999}; Alibert et al. \cite{alibert2005a}). We also
introduce a more elaborate treatment of photophoresis which is
applicable in all parts of protoplanetary disks and for all particle
sizes smaller than 10 cm.
 
%__________________________________________________________________

\section{Photophoresis}

The surface of a solid particle embedded in a gas is usually heated
heterogeneously by light. Gas molecules accommodated and rejected at
the surface will carry different momentum and a net force on the
particle results -- photophoresis (Rohatschek \cite{rohatschek1985};
Ehrenhaft \cite{ehrenhaft1918}). The effect is strongly pressure
dependent. It mostly goes unnoticed at normal atmospheric pressure on
Earth, but becomes strong for the pressures that are prevalent in
protoplanetary disks. Krauss \& Wurm (2005) and Wurm \& Krauss (2006)
used an approximation for photophoresis in the free molecular flow
regime. For small particles ($<1$ mm), this is applicable in regions
at radii $>0.5$ AU in the minimum mass solar nebula of Hayashi et
al. (\cite{hayashi1985}). However, the description for the denser
regions further in, and for larger particles further out, requires a
more detailed treatment of photophoresis. The conditions can be
described best by the Knudsen number, $Kn$, which is defined as
$Kn=l/a$, where $l$ is the mean free path of the gas molecules and $a$
is the radius of the particle in question. If the mean free path of
the gas molecules is no longer large compared to the particle sizes
considered, i.e. for $Kn<1$, the gas flow changes from free molecular
flow to the continuum regime. Under these conditions, the
photophoretic force also changes its pressure dependence from a linear
{\it{increase}} with pressure in the mean free molecular flow regime
to a linear {\it{decrease}} with pressure in the continuum
regime. This has to be taken into account. A semi-empirical approach
to a description of the photophoretic force $F_{ph}$ for all gas
pressures $p$ has been given by Rohatschek (\cite{rohatschek1995}). In
an alternative (and more rigorous) treatment by Beresnev et
al. (\cite{beresnev1993}) the photophoretic force on a spherical
particle, valid for both flow regimes, is given as

\begin{equation}
F_{ph}=\frac{\pi}{3}a^2IJ_1 \left(\frac{\pi m_g}{2kT} \right)^{1/2}
\frac{\alpha_E \psi_1}{\alpha_E+15\Lambda Kn(1-\alpha_E)/4+\alpha_E
\Lambda \psi_2},
\label{Fph}
\end{equation}

where $I$ is the light flux, $m_g$ is the average mass of the gas
molecules ($7.827 \times 10^{-27} \rm kg$), $T$ is the gas
temperature, and $k$ the Boltzmann constant. $J_1$ is the asymmetry
factor that contains the relevant information on the distribution of
heat sources over the particle's surface upon irradiation. In the
simplest case, when the incident light is completely absorbed on the
illuminated side of the particle, it is $J_1=0.5$, which we will
assume in the calculations below. The energy accommodation coefficient
$\alpha_E$ is the fraction of incident gas molecules that accommodate
to the local temperature on the particle surface and, thus, contribute
to the photophoretic effect. In the following, we assume complete
accommodation, i.e. $\alpha_E=1$. The corresponding coefficients,
$\alpha_n$ and $\alpha_\tau$, for the normal and tangential momentum
accommodation, respectively (Beresnev et al. \cite{beresnev1993}), are
already assumed to be 1, when expressing $F_{ph}$ by
Eq. (\ref{Fph}). The thermal relaxation properties of the particle are
summarized in the heat exchange parameter
$\Lambda=\lambda_{eff}/\lambda_g$, where $\lambda_g$ is the thermal
conductivity of the gas and $\lambda_{eff}$ the effective thermal
conductivity of the particle. This term includes the conduction of
heat through the particle, and the thermal emission from the
particle's surface, according to

\begin{equation}
\lambda_{eff} = \lambda_p + 4\epsilon \sigma T^3a.
\label{lambdaeff}
\end{equation}

Here, $\lambda_p$ is the thermal conductivity of the particle,
$\epsilon$ is the emissivity which we assume to be 1 further on, and
$\sigma$ is the Stefan-Boltzmann constant. The functions $\psi_1$ and
$\psi_2$ in Eq. (\ref{Fph}) depend only on $Kn$ in the form

\begin{equation}
\begin{array}{l}

\psi_1 = \frac{Kn}{Kn+(5\pi/18)} \left(1+\frac{2\pi^{1/2}Kn}{5Kn^2+\pi^{1/2}Kn+\pi/4} \right),

\\
\\
\psi_2 = \left(\frac{1}{2}+\frac{15}{4}Kn \right) \left(1-\frac{1.21\pi^{1/2}Kn}{100Kn^2+\pi/4} \right)

\label{psi12}
\end{array}
\end{equation}
 		
It should be noted that an additional photophoretic force arises if
the accommodation coefficients vary over the surface of the dust grain
(Cheremisin et al. \cite{cheremisin2005}), but we restrict the
treatment to the "classical" photophoretic force as given in
Eq. (\ref{Fph}). To summarize, Eq. (\ref{Fph}) is valid for all parts
of protoplanetary disks and all particles, and in the following we
assume $J_1=0.5$, $\alpha_E=1$, and $\epsilon=1$.

\section{The disk model}

In this paper, we use a rather simple model for the solar nebula,
namely an $\alpha$ 1+1D turbulent model. We provide a brief
description here, but more details can be found in Alibert et
al. (\cite{alibert2005a}), and Papaloizou \& Terquem
(\cite{papaloizou1999}).

In the $\alpha$-disk model, the radial evolution of the nebula is
governed by the following diffusion equation, modified to take
photo-evaporation into account:

\begin{equation}
{d \Sigma \over d t} = {3 \over r} {\partial \over \partial r } \left[
r^{1/2} {\partial \over \partial r} \nu \Sigma r^{1/2} \right] +
\dot{\Sigma}_w(r)
\label{eq_diff}
,
\end{equation}

where $\Sigma$ is the surface density of mass in the gas phase in the
nebula, $r$ is the distance to the Sun, and $\nu$ the mean (vertically
averaged) viscosity. The photo-evaporation term, $\dot{\Sigma}_w(r)$,
is taken to be the same as in Veras \& Armitage
(\cite{veras2004}). The mean viscosity is determined from the
calculation of the vertical structure of the nebula: for each radius,
$r$, the vertical structure is calculated by solving the equation for
hydrostatic equilibrium, together with the energy equation (which
states that the dissipation of energy by viscosity is balanced by the
removal of energy by diffusion), and the diffusion equation for the
radiative flux. The local viscosity (as opposed to that averaged in
the vertical direction) is calculated using the standard Shakura \&
Sunyaev (\cite{shakura1973}) $\alpha-$parametrization $\nu = \alpha
C_{s}^{2} /\Omega$, where the local speed of sound, $C_{s}$, is
determined by the equation of state, and $\Omega ^2 = G M_{\sun} /
r^3$. Using this procedure, we calculate, for each radius, $r$, from
the Sun, and for each value of the surface density, the distribution
of pressure, temperature and density. We then derive the mid plane
pressure and temperature, as well as the mean viscosity inside the
disk. These quantities are finally used to solve the diffusion
equation (Eq. (\ref{eq_diff})), and to calculate the pressure- and
temperature-dependant forces on dust grains as outlined in Sect. 4.

The structure of the nebula depends on two physical quantities, namely
the equation of state, and the opacity law. For the equation of state,
we use that developed by Saumon et al. (\cite{saumon1995}), taking the
fraction of Helium to equal 0.24. The opacity law used in our model is
taken from Bell \& Lin (\cite{bell1994}) at low temperatures, and
Alexander \& Ferguson (\cite{alexander1994}) at higher
temperatures. At low temperatures, the opacity law takes into account
the contributions from different types of grains, as well as their
sublimation, depending on the local temperature. Note, however, that
by modifying the local abundance of dust grains inside the nebula,
photophoresis may change the opacity, and therefore the structure of
the nebula itself. This retroaction is not taken into account in our
model, and will be the subject of future work.

\begin{figure}
 \centering
 \includegraphics[width=8cm, clip]{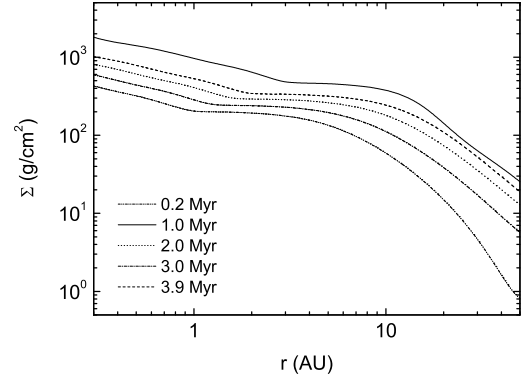}
 \caption{ The temporal evolution of the surface density of mass in
the gas phase as a function of the distance from the star.}
 \label{sigma}
\end{figure}

Some initial parameters have to be specified in order to calculate the
evolution of the solar nebula. We assume that the gas surface density
is initially given by a power law $\Sigma \propto r^{-3/2}$, with an
initial value taken to be $\Sigma (5.2 {\rm AU}) = 600 {\rm
g/cm}^2$. The mass of the nebula (between 0.25 AU and 30 AU) is $\sim
0.05 M_{\sun}$). The photo-evaporation rate is taken as being equal to
$\sim 1.5 \times 10^{-8} M_{\sun} / yr$, leading to a nebula lifetime
of the order of $\sim 3$ Myr. These values were used by Alibert et
al. (\cite{alibert2005c}) to calculate formation models of Jupiter and
Saturn consistent with the internal structure and atmospheric
compositions of these two planets. For the viscous
parameter, we set $\alpha = 0.002$. The radial dependence of the
resulting surface density at different ages of the disk is shown in
Fig. \ref{sigma}.

\section{Outward and inward drifts}

In a protoplanetary disk where the gas pressure (in the mid plane)
decreases with distance from the star, the gas is supported by a
pressure gradient, and rotates slower than the Keplerian velocity
(Weidenschilling \cite{weidenschilling1977}). Solid particles are only
stable on a Keplerian orbit. Therefore, interaction with the gas leads
to an inward drift of solids toward the star. For particles which
couple to the gas flow on timescales small compared to an orbital
period, the problem can be treated as being radial. The inward drift
is then induced by the fraction of gravity (residual gravity) which is
not balanced by the circular motion with the sub-Keplerian gas
velocity. The force, $F_{res}$, acting on a particle of mass $m_p$ due
to residual gravity is given as

\begin{equation}
F_{res}=\frac{m_p}{\rho_g}\frac{dp}{dr} .
\label{fres}
\end{equation}

Here, $\rho_g$ is the density and $p$ the pressure of the gas.
 
In addition, for small particles, radiation pressure also has to be
considered. Since we already assumed perfect absorption of particles,
we neglect the contribution to radiation pressure from scattering, and
obtain the following expression for the radiation pressure force

\begin{equation}
F_{rad}=\pi a^2 \frac{I}{c_{light}},
\label{frad}
\end{equation}
 
with $c_{light}$ being the speed of light. The sum of the outward
forces (Eq. (\ref{Fph}) and Eq. (\ref{frad})) and the inward force
(Eq. (\ref{fres})) is the drift force $F_{drift}$. We treat the
problem as being purely radial here, since we are mostly interested in
the small particles. These small particles couple to the gas on
timescales much smaller than the orbital timescale, which justifies
the radial treatment as outlined in Wurm \& Krauss
(\cite{wurm2006}). We estimate the radial drift velocity to be

\begin{equation}
v_{dr}=\frac{F_{ph}+F_{rad}+F_{res}}{m_p}\tau,
\label{vdrift}
\end{equation}

where $\tau$ is the gas grain coupling time. This section
considers a purely laminar disk, for which the drift of the particles
with respect to the gas {is adequately described by
Eq. \ref{vdrift}}. In the case of a turbulent disk, this is just
the average drift velocity, and turbulent diffusion must also
be considered. This will be treated in Sect. \ref{turbdrift}.

As larger dust aggregates drift outwards they pass from a region where
the continuum flow regime is valid to a region where the free
molecular flow regime applies. Therefore, as with the photophoretic
force, we have to consider an equation describing the gas grain
friction time in both regimes. It is

\begin{equation}
\tau=\frac{m_p}{6 \pi \eta a}C_c,
\label{tau}
\end{equation}

with $\eta$ being the dynamic viscosity of the gas. This assumes
Stokes friction, which is justified as the Reynolds numbers for the
drift of particles smaller than 10 cm are well below 1. The Cunningham
correction factor, $C_c$, accounts for the transition between the
different flow regimes (Cunningham \cite{cunningham1910}) and is given
as (Hutchins \cite{hutchins1995})

\begin{equation}
C_c = 1+Kn \left(1.231+0.47 e^{-1.178/Kn} \right).
\label{cc}
\end{equation}

For the calculation of $Kn$ in Eqs. (\ref{psi12}) and (\ref{cc}) we use the expression

\begin{equation}
l=\frac{2\eta}{\rho_g c}
\label{mfp}
\end{equation}

for the mean free path , $l$, of the gas molecules (Beresnev et
al. \cite{beresnev1993}), where $c$ is the mean thermal velocity of
the gas molecules. For large Knudsen numbers, the combination of
Eqs. (\ref{tau})-(\ref{mfp}) yields

\begin{equation}
\tau_{Ep}=\varepsilon \frac{m_p}{\sigma_p} \frac{1}{\rho_g c},
\label{tau_epstein}
\end{equation}

where $\sigma_p$ is the geometrical cross section of the dust
particle. This expression corresponds to the stopping time in the
Epstein friction regime as given by Blum et al. (\cite{blum1996}). The
empirical parameter $\varepsilon$ has a value of $0.68 \pm 0.10$.

The dynamic viscosity $\eta$ is directly related to the square root of the gas temperature, $T$. Thus,

\begin{equation}
\eta = \eta_{0}\sqrt{T/T_{0}}
\label{eta}
\end{equation}

with $\eta_{0}=8.4 \times 10^{-6} \rm Pas$ at $T_{0}=280 \rm K$. 

\begin{figure}
 \centering
 \includegraphics[width=8cm, clip]{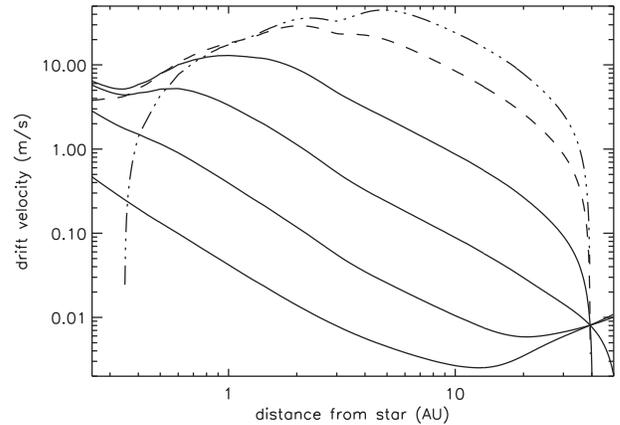}
 \caption{ Outward drift velocity of particles as a function of the
radial distance from the star, if a disk of age 0.2 Myr were
completely transparent. The curves represent particles of different
sizes. From bottom to top, the solid curves denote particles of $a=1,
10, 100 \rm \mu m$, and $a=1 \rm mm$. The dashed lines are for
particles of $a=1 \rm cm$ and $a=3 \rm cm$. The value for 3cm
particles turns negative inside approximately 0.4 AU. For particles
larger than 1mm the drift velocity also becomes negative beyond ~40
AU.}
 \label{vdrplot}%
\end{figure}

For the calculations, we assume the bulk density of the dust particles
to be $\rho_p=1 \rm g/cm^3$, which is a reasonable value for dust
aggregates. For the thermal conductivity of the dust particles, we
take $\lambda_p=0.01 \rm W/mK$. This can also be taken as a reasonable
assumption for dust aggregates (Presley \& Christensen
\cite{presley1997}), but might be subject to debate in the future. In
addition, any differences in the properties of these particles might
result in some subtle selection effects. For the thermal conductivity
of the gas, $\lambda_g$, we adopt values for molecular hydrogen for
temperatures above 150K (as tabulated by Incropera \& DeWitt
(\cite{incropera2002})). For smaller temperatures, helium has a larger
thermal conductivity than hydrogen. With a significant fraction of the
gas being helium, we assume that this helium will then determine the
thermal conductivity of the gas. We use values for helium taken from
the compilation of Bich et al. (\cite{bich1990}).  The radial
variation of the drift velocity according to Eq. (\ref{vdrift}) is
plotted in Fig. \ref{vdrplot} for various particle sizes. This shows
how particles would drift in a transparent model. The values are only
to be taken as valid for particles in the inner cleared region, out to
the optically thick edge of the disk.

Positive values of $v_{dr}$ correspond to an outward drift, while
negative values denote an inward drift (not plotted on this
logarithmic plot). It should be pointed out that these drift
velocities are with respect to the radial motion of the gas. The
resulting velocities with respect to the central star might turn out
to be negative, even for positive values of $v_{dr}$, when the inward
gas motion due to stellar accretion is taken into account. This will
be considered in Sect. 5 for the temporal evolution of the dust
distribution. The first thing to note is that points of equilibrium
occur where the drift vanishes. Stable equilibrium points where
particles become concentrated are described in Krauss \& Wurm
(\cite{krauss2005}) and Wurm \& Krauss (\cite{wurm2006}). In the disk
model used here, at a time of 0.2 Myr, these stable points can be
found at a distance of 40 AU for particles larger than 1mm. Particles
are eventually pushed outwards to this concentration
distance. Particles smaller than about 10 $\rm \mu m$ are pushed
further outwards due to radiation pressure. However, photophoresis has
two branches. If the mean free path of the gas molecules is smaller
than the particle size, the photophoretic force decreases with
pressure. This is responsible for a second equilibrium, which lies
closer to the star for somewhat larger particles. This is seen in
Fig. \ref{vdrplot} for 3 cm particles which drift inwards closer than
$\approx$ 0.4 AU from the star. This is not a stable equilibrium, and
the particles eventually move to one side or the other of this
point. Nevertheless, particles might stay in this region for a
prolonged time, long enough to be subject to other processes, such as
collisions with planetesimals. For larger particles, the location of
the maximum drift velocity shifts to larger distances from the
star. For moderate sizes, from micrometer up to a few cm, the
magnitude of the drift velocity increases with particle size
everywhere in the disk. Therefore, the important point is that the
slowest moving particles are usually the smallest. For objects larger
than about 1m, the radial treatment would no longer hold since the gas
grain friction times become comparable to the orbital period. Also,
rotation becomes important for such large bodies, which will change
the photophoretic force. We therefore restrict our results to
particles smaller than about 10cm, which we will call dust.

Starting with a disk which has a transparent inner region and an
opaque outer region, dust particles at the inner edge are subject to
photophoresis. Large particles move faster than small
particles. Therefore, they are rapidly moving outwards. The slower,
smaller particles remain behind and the larger particles are soon
shielded from starlight. Therefore, the edge moves outwards with a
drift velocity determined by the smallest particles that are able to
render the disk optically thick. As the larger particles react
swiftly, the edge is rather sharp, as observed for the transient disks
mentioned above.

\subsection{\label{viscous}Accretion due to viscous evolution}

It has to be remembered that, in addition to the relative
motion between solids and gas we must also consider the absolute
motion of the gas. As the disk evolves viscously, it accretes onto the
star with an accretion speed $v_{ac}$. Since the accretion
speed depends on the evolutionary stage of the disk (in other
words, on the elapsed time), it can act to delay the outward motion
of the dust particles and  then, for a short period, act to slow
it down. For the young disk in the model considered here, the
accretion speed at 1 AU is about 20 $\rm cm s^{-1}$, dropping to 1
$\rm cm s^{-1}$ beyond 10 AU. At later times, the inner value drops to
a few $\rm cm s^{-1}$ while the value at larger distances changes
little. As can be seen in Fig. \ref{vdrplot}, the accretion is
comparable to the outward drift for 1 $\rm \mu m$ particles with an
assumed thermal conductivity of $\lambda_p = 0.01$ $\rm W m^{-1}
K^{-1}$, but as soon as the particles reach 10 $\rm \mu m$ or larger,
or if they have a lower thermal conductivity, accretion ceases to be
an obstacle to outward drift.

\subsection{Turbulent diffusion\label{turbdrift}}

In a turbulent disk, a fraction of the particles experience a
faster or slower inward drift due to turbulent diffusion. This
acts to shift the transition region between the optically thick
outer disk and the optically thin inner region further inwards. If
this is more effective than photophoresis, no outward drift of the
inner edge will occur. If the inner edge diffuses inwards 
more slowly than it moves outwards due to photophoresis, then
diffusion will not completely stop the outward motion, but will
slow down or change the characteristics of the clearing of solids from
the disk.

The fraction of diffusing particles that is necessary to
move the transition between transparent and opaque disk inwards
depends on the particle sizes and number densities present at a given
time and location, together with their diffusion speeds.  A
detailed treatment of turbulent diffusion working together with
photophoretic drift is beyond the scope of this paper and will
be addressed in detail in future studies.  However, some
reasonable and some extreme estimates for turbulent diffusion are
given here to illustrate the effect of turbulence on the photophoretic
evolution of a disk. The case in which diffusive drift is
underestimated to the greatest extent has already been given by the
laminar case, which assumes that a particle which is picked up
by a turbulent eddy and is transported inwards will be transported
outwards again by the same eddy, which will return it to
its original position. No net drift would result. An extreme
overestimation is case where particles are handed inwards from
one of the large eddies to another large eddy, and so on, with a
resulting maximum speed of the turbulent gas of more than 100 m/s
(Supulver \& Lin (\cite{supulver2000}); Dullemond \& Dominik
(\cite{dullemond2000})). If that were the case for a large fraction of
particles photophoresis would not be able to clear an optically thick
disk at all as photophoretic drift velocities are, in general, smaller
than 100 m/s. However, turbulence can be described as a cascade of
eddies of different size, and inward as well as outward motions occur
as a particle adapts to the changing gaseous environment. Therefore,
the effective diffusion speed is somewhere between the extreme
cases. This kind of random walk in a turbulent cascade has 
already been studied (for example, by Supulver \& Lin
(\cite{supulver2000})).

For simplicity, we assume here that turbulent diffusion does not
depend on the particle size for the range of particle sizes considered
($ < $ 10 cm), and that inward diffusion is described by a constant
velocity, $v_{turb}$.  Certainly, in general, this is misleading
as diffusion cannot be defined as a velocity. Diffusion is
typically characterized by the distance travelled by a particle 
as a function of the square root of time. However, keeping that
in mind, it suffices for estimating the possible effects in
this case.

\begin{figure}
 \centering
 \includegraphics[width=8cm, clip]{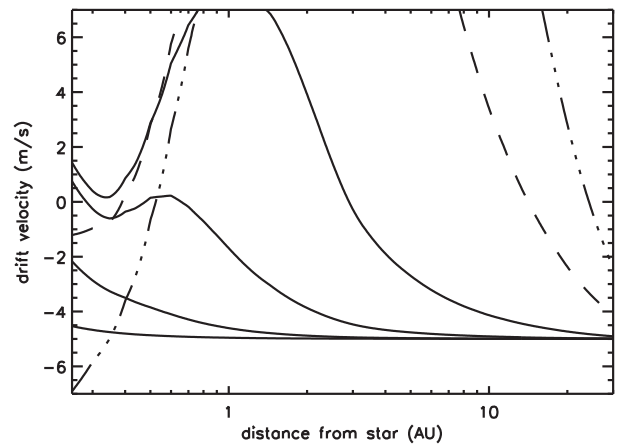}
 \caption{Particle drift and turbulent diffusion at 0.2 Myr.  As
in Fig. \ref{vdrplot} from bottom to top, the solid curves denote
particles of $a=1, 10, 100 \rm \mu m$, and $a=1 \rm mm$. The dashed
and dash-dotted line are for particles of $a=1 \rm cm$ and $a=3 \rm
cm$, respectively. A constant turbulent inward diffusion of 5 m/s has
been subtracted for all particle sizes.}
 \label{vdrplotdiff}
\end{figure}

Realistic values of a turbulent diffusive drift should be
significantly larger than the laminar case ( $v_{turb} = 0 \rm
ms^{-1}$ ), but also much smaller than the extreme ( $v_{turb}
\ll 100 ms^{-1}$ ).  Detailed particle tracks in a turbulent disk are
given in the Monte Carlo simulations by Supulver \& Lin
(\cite{supulver2000}). One of their important results is that 85\% of
5 mm particles placed between 2 and 3 AU move outwards rather
than inwards. However, the detailed track of a 1 cm particle shows
numerous parts of its trajectory where it moves inwards with a maximum
speed of about 5 m/s (estimated from their Fig. 8b). Supulver \& Lin
(\cite{supulver2000}) assume a viscous parameter $\alpha = 10^{-2}$,
which is comparable to the value of $\alpha = 0.002$ assumed in our
model. Therefore, a maximum turbulent diffusion drift for the
optically thick edge in our model might be estimated as $v_{turb} = 5
\rm m/s$. It should be noted that photophoretic drift is counteracted
by a much smaller diffusive drift if the fast diffusing particles are
not numerous enough to increase the opacity of the disk. More
realistic values might therefore be much smaller than 5 m/s, but a
detailed treatment is postponed to future studies of diffusion at the
inner edge.  In any case, diffusion will slow down the photophoretic
motion of particles. Fig. \ref{vdrplotdiff} illustrates what effect
diffusion imposes on particles of different size. The drift velocities
according to Eq. (\ref{vdrift}) are used, but a constant diffusive
drift of 5 m/s is subtracted. In a disk (stage) that has an opaque
outer region and that is dominated by particles with
negative drift values, no outward drift is possible. In our model,
this is the case if the turbulent disk is dominated by particles
smaller then 100 $\rm \mu m$. This result is not very sensitive to the
exact value of the diffusion speed of 5 m/s. As small particles 
already drift rather slowly in the laminar case, the diffusion speed
has to be a factor of 10 smaller before being considered
negligable. The existence of a sufficient fraction of particles
smaller than $100 \rm \mu m$ might therefore offset any clearing of
the disk until either the turbulence subdues significantly or
the small particles vanish (through, for example, coagulation).

In addition to these considerations, cm-sized particles cannot
move away from a close inner rim (at about 0.5 AU or less) at early
times as the photophoretic force acting on them is weak due to
the small Knudsen numbers. An initial early clearing is only
possible if a monodisperse size distribution of particles somewhat
smaller than 1 mm exists. Once a gap of 0.5 AU is formed (one way
or the other), or if the gas density in the inner region decreases,
cm-size particles can also be pushed further out. It is
interesting to note that in a turbulent disk consisting of mm-sized
dust particles, the inner edge might not move further out than a few
AU (3 AU with the given parameters). The position of the inner edge
will then be almost constant even as the disk evolves and the gas
density decreases, which will be shown in Sect. \ref{turbmotion}.

In a laminar transparent disk, particles become concentrated 
at more or less the same distance (several tens of AU)
independent of their size. Caution should be used when
comparing this scenario with the particle size dependent position of
the inner edge.  The same size independent concentrations at large
distances will also occur in a turbulent disk if the disk is
completely transparent.  The position of the edge is merely
smeared out by turbulence but is not shifted by diffusion. A size
dependent position of the inner edge requires the outer disk to be
opaque so that particles can be illuminated or shadowed.

\section{\label{temp}Temporal evolution of the dusty edge}

In the model of photophoretic clearing discussed here, the smallest
particle size is the critical parameter. Unfortunately, little can be
said, in general, about the evolution of the size of the smallest
particles abundant enough to make the disk optically thick.

In simple growth models that only consider sticking of particles in
collisions, the growth is incredibly quick.  Particles will grow to
cm-size in a few thousand years (Dominik et al. \cite{dominik};
Weidenschilling \cite{weidenschilling2000}). This simple growth
mechanism to cm-size is in full agreement with experimental findings
(Blum et al. \cite{blum2000}; Blum \& Wurm \cite{blum2000b}). However,
as soon as particles grow further, collisions will start to produce a
new population of smaller particles. Recent experiments show that the
growth of a larger body and the generation of smaller fragments of
dusty bodies can occur in the same collision (Wurm et
al. \cite{wurm2005}). This will replenish the reservoir of small
particles but, unfortunately, there is currently no realistic
prediction for the size distribution of small particles resulting from
fragmentation and growth, and, due to the lack of more reliable data,
we will assume (somewhat arbitrarily) an exponential evolution here,

\begin{equation}
a=a_0 e^{t/t_0}
\label{size}
\end{equation}

with $a_0$ and $t_0$ being free parameters. It is plausible that, as
larger bodies form and the amount of matter redistributed to the dust
component gets smaller, growth by coagulation will dominate, and the
smallest particle size will shift to larger values. Otherwise, the
details might be responsible for the differences in the evolution of
different systems (for example, the position of the inner dust edge at
different distances for stars of similar ages). Once the minimum
particle size has reached a value of 1cm, we turn the temporal
evolution of particle size off. Thereafter, we calculate particle
drifts assuming a constant size of 1 cm. This accounts for the fact
that, for reasonable nebula masses, particles larger than this cease
to have a significant effect on the opacity of the disk. It is also
tied into the fact that our treatment of photophoresis only applies to
particles smaller than $10 \rm cm$. Particles of cm-size move so
rapidly that they essentially trace the position of their equilibrium
point where all radial forces are balanced. This equilibrium exists
for particles larger than 1 mm and its position is independent of
particle size (see Fig. \ref{vdrplot}).  Therefore, the cm-sized
particles effectively trace the position within the disk where the
small particles that now make up an optically thin fraction of the
disk will become concentrated. These particles represent a combination
of replenished material and the remains of the initial dust
population, and are between 1mm and 10cm in size.

\subsection{\bf{Particle motion in a laminar disk}}

With the assumptions and disk model given above, we investigate the
motion of the edge due to the effect of photophoresis acting on the
smallest particles.  Therefore, the position of the edge,
$R(t_{disk})$, at the age of the disk, $t_{disk}$, is calculated by
numerically solving

\begin{equation}
R(t_{disk})=\int_{0}^{t_{disk}} (v_{dr}(a(t),R(t))-v_{ac}(R(t))  dt.
\label{integral}
\end{equation}

It has to be noted that the assumptions above do not
consider the diffusion of particles due to turbulence (as
discussed in Sect. \ref{turbdrift}). Only the average particle motion
due to the viscous evolution of the disk is taken into account by
including $v_{ac}$. As outlined above, a fraction of particles which
diffuse inwards faster than average due to turbulent diffusion has the
potential of rendering the disk optically thick. We discuss the
possible effects of turbulent particle transport on the evolution of
the inner edge in Sect. \ref{turbmotion}.

\begin{figure}
 \centering
 \includegraphics[width=8cm, clip]{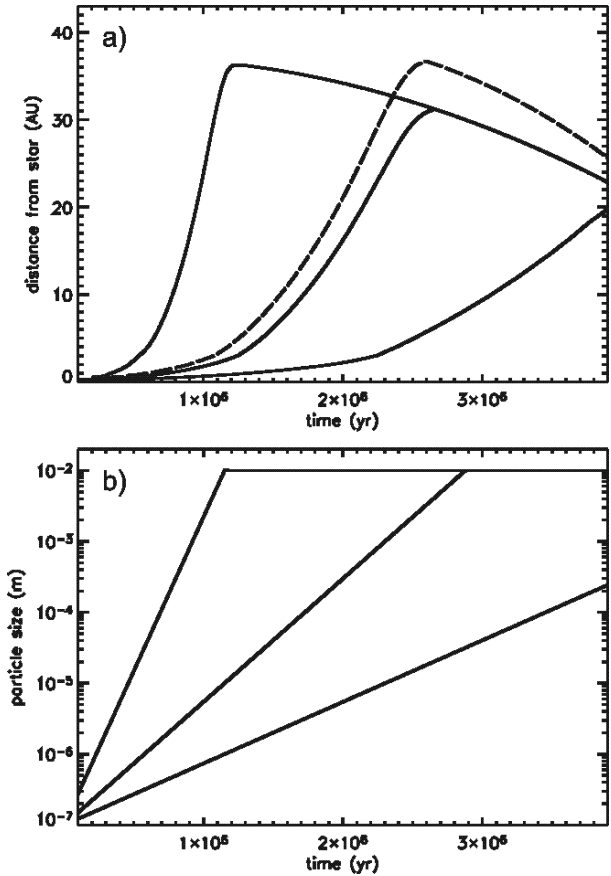}
 \caption{ a) The position of the inner edge of the dust disk as a
function of time for different particle growth timescales. For the
solid lines, the growth timescale is $t_0 = 100,000$ yr, $t_0 =
250,000$ yr, and $t_0 = 500,000$ yr (moving from left to right), and
$\lambda_p = 0.01$ $\rm W m^{-1} K^{-1}$. The dashed line is for $t_0
= 250,000$ yr and $\lambda_p = 0.005$ $\rm W m^{-1} K^{-1}$. The
evolution is divided into 3 parts. With increasing time, these parts
correspond to slow drift, fast drift, and an inward retreat.  b)
Corresponding particle size evolution. The growth is cutoff once the
particles have reached a size of 1cm. Comparison with a) shows that
the edge only starts to move outwards efficiently once particles have
grown to $10 \rm \mu m$ or larger.}
 \label{theedge}
\end{figure}

We start with particles placed at the inner edge of the model disk, at
0.25 AU, which is in good agreement with the initial sublimation
distance since the temperature is much larger than 1000K at this
location in our model disk. We assume that the particles are initially
of interstellar size ($a_0=0.1$ $\rm \mu m$). As our timescales for
growth, we take $t_0 = 100,000$ yr, $t_0 = 250,000$ yr, and $t_0 =
500,000$ yr. We also calculate the evolution of particles with a
growth timescale of $t_0 = 250,000$ yr, but a lower thermal
conductivity of $\lambda_p = 0.005$ $\rm W m^{-1} K^{-1}$. The
evolution of the position of the inner disk edge for these four
scenarios is shown in Fig. \ref{theedge}a. The underlying growth of
the particles is depicted in Fig \ref{theedge}b. As can be seen, the
edge starts to move outwards after a few 100 kyr, as the accretion
becomes slow enough, and the particles grow sufficiently to move more
rapidly.

The evolution of the dusty edge can be divided into three discrete
phases. For the first few 100kyr, the edge does not move at all, or
only moves very slowly due to the combination of the high accretion
rate and the small particle sizes. As can be seen in
Fig. \ref{theedge}b, this corresponds to a period when the
particles are smaller than $10 \rm \mu m$. Once the particles have
grown beyond that size, the dusty edge moves outwards. When the
particles have grown larger than about 1 mm they rapidly move to the
equilibrium position, where all drift forces are balanced. As growth
proceeds over time, it is likely that the overall dust density at the
edge, and in the optically thick outer part, decreases strongly with
time. In fact, as our model assumes growth of the particles, it is
conceivable that the whole disk might turn optically thin before it
reaches the end of its viscous evolution. In this case, the edge at
later times would no longer be an inner edge but rather a
concentration of the remaining or replenished particles in a ring-like
distribution, as described in Krauss \& Wurm (\cite{krauss2005}) and
Petit et al. (\cite{petit2006}). As the gas density in the disk
decreases over time, the photophoretic force on the dust grains gets
weaker, and the equilibrium position moves inward again leading to a
dust ring with decreasing radius (Petit et al. \cite{petit2006}). The
length of the three evolutionary phases is obviously directly
connected to the time constant of the particle growth process. The
faster the particles grow, the earlier the edge starts to move outward
and reaches its maximum distance from the star. Furthermore, the
maximum radius that is reached by the edge or the ring depends on how
fast the dust particles grow (and also on the thermal conductivity of
the grains).

\subsection{\label{turbmotion}Particle motion in a turbulent disk}

\begin{figure}
 \centering
 \includegraphics[width=8cm, clip]{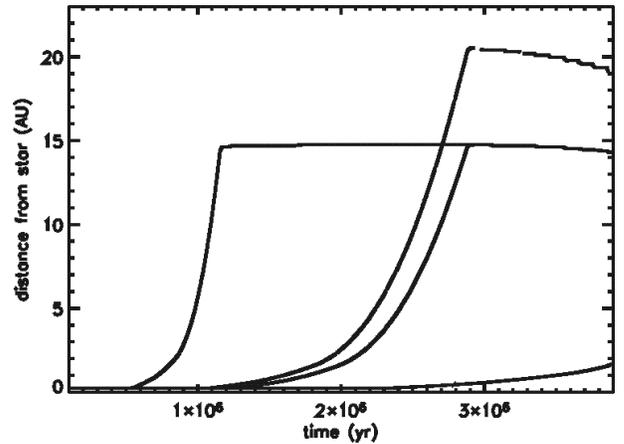}
 \caption{\bf{Same as Fig. \ref{theedge}a but assuming 5 m/s offset due to turbulent diffusion.}}
 \label{theedge2}
\end{figure}

The same calculations as those described in the previous
subsection are carried out assuming that turbulent diffusion would
counteract the photophoretic outward drift by a constant offset of 5
m/s. The resulting radial velocities are shown in Fig. \ref{theedge2}
with the corresponding scenarios for the particle size evolution being
the same as in Fig. \ref{theedge}b. As discussed in
Sect. \ref{turbdrift}, small particles cannot oppose turbulent
diffusion. Particles have to grow larger than $100 \rm \mu m$ for
photophoresis to start clearing the disk. As a consequence of this,
the initiation of the outward drift of the edge is delayed
longer than in the case of a non-turbulent disk. Nevertheless,
the first two phases of the disk evolution are qualitatively similar
to the laminar case, since the photophoretic drift of small particles
is very slow and does not contribute significantly to the clearing
process. However, the maximum distance achieved from the star and
the behaviour during the third evolutionary phase, which is
characterised by an inward motion of the edge in the laminar case, are
considerably different in a turbulent disk. Due to the additional
inward velocity ($v_{turb}$) the edge reaches a maximum radius
of only about 15AU for $\lambda_p = 0.01$ $\rm W m^{-1} K^{-1}$ and
20AU for $\lambda_p = 0.005$ $\rm W m^{-1} K^{-1}$. Once the edge has
reached this maximum distance it stays there, and barely moves
while the gas is dispersed. This is a result of the fact that, if
the photophoretic drift relative to the gas dominates the particle
motion, the drift velocity is almost independent of the absolute gas
density (Wurm \& Krauss \cite{wurm2006}). For the assumed value of
turbulent diffusion, the net drift velocity equals the
outward drift velocity when it is solely dominated by the
photophoretic term, and is not affected by residual
gravity. This is in contrast to the gas density dependent position
where all drift forces are equal.  Thus, the disk evolution has
little influence on the position of the inner edge until the edge 
approaches the location of its inward moving laminar counterpart.

For the largest particles (cm-sized), however, the offset of 5
$\rm ms^{-1}$ might not be justified if the bulk of the disk is already
transparent. In that case, turbulent diffusion would only smear out the
distribution around the stable position of the concentration points
given by the laminar case. This is, in a general sense, true for any
transparent disk. Even small particles are then concentrated at their
equilibrium distance as determined for the laminar case since
turbulence no longer has the ability to move particles from the
illuminated region into the shadow.

\begin{figure}
 \centering
 \includegraphics[width=8cm, clip]{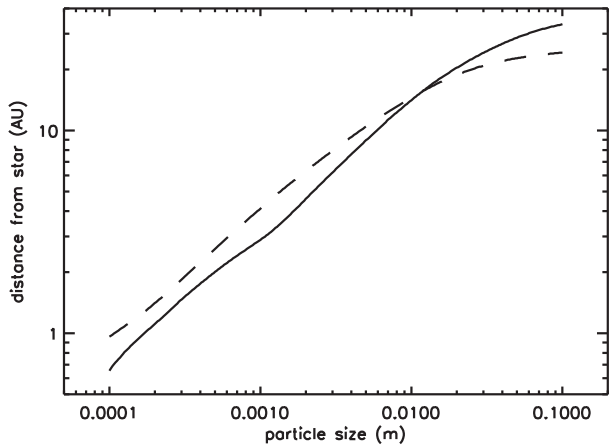}
 \caption{\bf{Equilibrium position as a function of the minimum particle size for a turbulent disk
at the age of 0.2Myr (solid line) and 3.4Myr (dashed line).}}
 \label{sizedependence}
\end{figure}

The turbulent offset leads to a dependence of the equilibrium position
of the inner edge on particle size. If we do not cut off the particle
size at $a = 1 \rm cm$, but at different sizes, the position of the
inner edge shifts. In equilibrium, the inner edge is where the
drift velocities turn negative in Fig. \ref{vdrplotdiff}. These
positions are plotted in Fig. \ref{sizedependence} as a function of
the particle size at a disk age of 0.2Myr (solid line) and 3.4Myr
(dashed line). The dependence on the particle size is obvious. It can
also be seen that the position of the edge changes very little during
more than 3Myr of disk evolution (for a given particle size). For
particle sizes $<1cm$ it is shifted slightly outward, whereas for
larger particles it moves slightly inwards.

Therefore, the qualitative differences between the laminar
and the turbulent cases can be summarised as follows. When
turbulent diffusion is active, the clearing of the inner disk
begins later, the inner edge does not move as far out, and the
position of the edge depends much more on the size evolution of the
particles than on the dispersion of the gas component of the disk.

\section{Possible applications}

The evolution of disk material described in this work lends itself to
a number of possible applications. A detailed discussion of these is
beyond the scope of this paper, and will be carried out in a later
work. However, a few simple examples serve to highlight the ideas and
predictions which could be based upon photophoretic disk clearance.

In our model, the motion of the inner edge of the disk over time 
in the laminar case is strongly dependent on the assumptions made
about the evolution of the dust particles. A different growth rate to
that assumed in Eq. (\ref{size}) can lead to significant changes in
each of the three regions described above, either leading to an
initial enhancement in the clearing rate, or acting to prevent the
formation of a larger hole, if the disk disperses more rapidly than
the smallest particles evolve. Such a wide range of possible outcomes
is in close agreement with the diversity of transient disks found
(Calvet et al. \cite{calvet2002}; Forrest et al. \cite{forrest2004};
D'Alessio et al. \cite{dalessio2005}; Bergin et
al. \cite{bergin2004}).  If turbulent diffusion is important, the
size of the inner hole of transient disks might be determined by the
cut-off particle size rather than the growth rate and, therefore, be
less dependent on time.

The formation of dust rings (Krauss \& Wurm \cite{krauss2005}; Petit
et al. \cite{petit2006}) around youthful stars can be well explained
using this model. These systems can be taken as representing old
disks, nearing the end of their photophoretic evolution. If dust
particles grow in such a system, this will lead to all the material
eventually being devoured, and the rings will no longer be dominated
by dust that evolved at the very location of the edge/ring. Dust
fragments which are produced at smaller radii than the ring location
(such as those created in collisions (Wurm et al. \cite{wurm2005}) or
coming from eolian erosion (Paraskov et al. \cite{paraskov2006}))
would be pushed rapidly outward, and would represent a significant
contribution to the ongoing survival of the ring. The equilibrium
positions might thus become more and more dominated by processed dust
from the inner regions. As the material concentrates in the regions
where comets were supposedly formed (see Horner et
al. \cite{horner2006}, for a discussion of the different plausible
formation regions for cometary bodies), this might prove to be an
important mechanism in explaining the abundance of unexpected material
within these comets, such as the presence of crystalline silicates or
high temperature minerals which have been detected in different
objects (Campins \& Ryan \cite{campins1989}; Bouwman et
al. \cite{bouwman2003}; Sitko et al. \cite{sitko2004}).
 
The timing of the formation and motion of the ring when compared to
the formation and ejection epochs of the comets would definitely be an
area of interest. Depending on the formation model accepted for the
Oort cloud and Edgeworth-Kuiper belt, it is possible that the
photophoretic dust loading would lead to observable differences
between objects which reside in different reservoirs (for example, one
might expect the bulk of comets that have resided in the
Edgeworth-Kuiper belt to have experienced significant dust loading,
while objects ejected into the Oort cloud may have escaped prior to
the passage of the ring, leading to a much lower concentration of dust
and high-temperature materials). Even at the most basic level, it may
be that the least dusty comets represent those that were ejected to
the Oort cloud at the earliest epoch, before the dusty edge had time
to sweep their formation region.

As the edge moves outward and collects particles smaller than a few cm
in size, it creates a region with a strong density enhancement. It is
very likely that the edge could act as a trigger or boost of
planetesimal formation, either by gravitational instabilities or
through the mantling of preexisting planetesimal cores with an
enhanced flux of small particles.

The case of the regular and irregular satellites of the gas and ice
giant planets represents another possible application for this
model. The regular satellites of these planets most likely formed
along with the planets themselves (Alibert et
al. \cite{alibert2005d}), from the same material at the same time. The
irregular satellites, by contrast, likely represent a population of
captured bodies which formed elsewhere in the Solar system ({\'C}uk \&
Gladman \cite{cuk2006}, Jewitt \& Sheppard \cite{jewitt2005}). It is
therefore plausible that the irregular satellites may contain
significantly different concentrations of dust and high-temperature
materials when compared with the regular satellite families.

Some authors (Ida et al. \cite{ida2000}) have suggested that the Solar
nebula was truncated by the passage of a nearby star during the latter
stages of planet formation. Such a mechanism has been invoked to
explain the apparent sharp outer edge of the Edgeworth-Kuiper belt,
and the presence of Sedna-like objects (Kenyon \& Bromley
\cite{kenyon2004}). If such an encounter did occur, it is quite
possible that it could have affected the concentration of dust (which
in the late stages of planet formation would be expected to be at
$\sim$ 30 or 40 AU), perhaps even helping to strip this material from
our Solar system entirely, aiding the truncation and diminution of the
disk itself.

Finally, it is even possible that the enhanced density present in the
outward-moving ring would help enhance the rates of chemical reactions
in the solar nebula, through the increased concentration of catalytic
material. Such enhanced reaction rates could play an important role in
modifying the chemistry of the region in which the giant planets
formed. In particular, the supply of high temperature material (in the
form of iron or nickel grains) may ease the conversion of CO into
CH$_4$ in the outer solar nebula via Fischer-Tropsch catalysis (CO +
3H$_2$ $\leftrightarrows$ CH$_4$ + H$_2$O), since the efficiency of
this reaction increases with the growing concentration of metal
particles (Sekine et al. \cite{sekine2005}). The production of CH$_4$
in the outer solar nebula in this manner has already been discussed in
the literature (Prinn \& Fegley \cite{prinn1989}; Kress \& Tielens
\cite{kress2001}), but the overall rate of CH$_4$ formation is
difficult to quantify because the total amount of available catalyst
surface is unknown (Kress \& Tielens \cite{kress2001}). On the other
hand, since homogeneous gas-phase chemistry predicts that CO is the
dominant carbon species in the solar nebula (Mousis et
al. \cite{mousis2002}), an increase in the concentration of metallic
grains in the zone of giant planet formation, resulting from the
outward transit of a dust ring, could lead to an efficient production
of CH$_4$. This fits well with the current composition of Titan's
atmosphere, where CH$_4$ is the dominant carbon compound.

Moreover, the enrichments of volatiles measured in the atmospheres of
Jupiter and Saturn (Owen et al. \cite{owen1999}; Flasar et
al. \cite{flasar2005}) have recently been interpreted as requiring an
over-solar abundance of water in order to facilitate the trapping of
volatiles as clathrate hydrates within the planetesimals produced in
the outer nebula (Alibert et al. \cite{alibert2005b}; Mousis et
al. \cite{mousis2006}). Such an increase in the water concentration
inside the zone of giant planet formation is still poorly
understood. Therefore, since Fischer-Tropsch catalysis has also been
suggested as an efficient means of producing H$_2$O in astrophysical
environments (Willacy \cite{willacy2004}), an enhancement of the
density of metal particles in this region could explain the presence
of large amounts of H$_2$O.

\section{Particle rotation}

The first idea to apply photophoresis to the behaviour of
protoplanetary disks is still only one year old (Krauss \& Wurm
\cite{krauss2005}). Since then, doubts have been raised on a number of
occasions as to the general applicability of photophoresis to dust
grains if the particles rotate. In order to answer these concerns, we
present the following comprehensive summary of arguments that show
that the rotation of dust particles which would be expected in a
protoplanetary disk does not prevent their photophoretic drift
described above. In previous work, some of these arguments have been
mentioned in passing (Krauss \& Wurm \cite{krauss2005}, Wurm \& Krauss
\cite{wurm2006}), but this is the first time that they have been
covered in detail.

The most important point to keep in mind in this context is that we
deal with a gaseous disk where any motion of small particles, both
linear and rotational, is quickly damped. This behaviour stands in
stark contrast to that of the particles in the current Solar system,
which retain any rotation for a long period of time. In general, to
prevent photophoretic drift, the particles must fulfill both of the
following criteria :

\begin{itemize}
\item the rotation must be such that the illuminated rotates into
shadow, and the shaded side into sunlight, and
\item the rotation period of a particle must be shorter than the time
taken to reorientate the temperature gradient across its surface.
\end{itemize}

The timescale for thermal relaxation is established by the conduction
of heat through the particle, the thermal emission from the particle
surface, and the transfer of heat to colliding gas molecules. Which of
these modes dominates the system depends on the particle size and on
its thermal properties, the temperature, and the gas density.
However, it is important to note that all this becomes unimportant if
the rotation axis is oriented in radial direction, lying parallel to
the direction of incident radiation. In that case, the temperature
gradient will be fixed in space, and photophoresis can act
undisturbed, independent of the rotation speed.

Dust particles achieve thermal rotation rates of
$\omega_T=(2kT/I_p)^{1/2}$ (where $I_p$ is the particle's moment of
inertia), around a randomly oriented axis. Krauss \& Wurm
(\cite{krauss2005}) showed that this rotation rate is significantly
slower than the thermal relaxation timescale from heat conduction, for
typical particle properties.  By following the same formalism as
Krauss \& Wurm (\cite{krauss2005}), it can be estimated that, for the
particle sizes and thermal conductivities considered here, the thermal
relaxation time due to heat conduction is one to four orders of
magnitude shorter than the time constant, for a thermally induced
$180^o$ rotation.

Collisions between dust particles will excite rotations about
arbitrary axes. To examine the effect of particle-particle collisions,
we compare the mean time span between two collisions experienced by a
particle with the stopping time $\tau$ as defined in
Eq. (\ref{tau}), which we assume to be of the same order of
magnitude as the rotational stopping time. The similarity might be
based on the following analogy. A very large but thin plate moving
through a viscous medium with the thin section ahead will be
decelerated by the frictional interaction between the two sides and the medium
(often used to define viscosity). If the plate is bent to form a
hollow cylinder, and made to rotate, it is the same
interaction on the outside and inside of the cylinder which decelerates the
rotation. Thus, the translational motion of the plate and the rotational motion
of the cylinder are stopped
on the same timescale. In detail, the exact ratio between the stopping
time for rotation and for linear motion depends on the morphology of
the particle of interest. However, the coupling mechanisms are the
same, and for simplicity we estimate the stopping time for rotation
according to Eq. (\ref{tau}). The collision time is given by

\begin{equation}
\tau_{coll}=(n_p v\sigma_{coll})^{-1}
\label{taucoll}
\end{equation}

where $n_p$ is the particle number density, $v$ is the relative
velocity of the colliding particles, and $\sigma_{coll}$ is their
collisional cross section. For the particle number density $n_p$, we
assume a dust to gas mass ratio of 1\%, and that all of the mass is
contained in particles of radius $a$, which gives

\begin{equation}
\tau_{coll}=33 \frac{\rho_p a}{\rho_g v}.
\label{taucoll2}
\end{equation}

The ratio of the two time constants for collisions and coupling is then

\begin{equation}
\frac{\tau_{coll}}{\tau}=\frac{150 \eta}{\rho_{g} v a C_c}.
\label{ratio}
\end{equation}

For all relevant combinations of parameters, this ratio is much larger
than 1. A population of 1 cm particles, for instance, with a density
of $1 \rm g/cm^3$ at 1AU in a 0.2 Myr old disk yields
$\tau_{coll}/\tau=1.23 \times 10^6$, assuming a relative velocity of
$v = 1 \rm ms^{-1}$ between the individual particles. If all particles
were $a=1 \rm \mu m$ in size, the collision velocities would be solely
thermal, on the order of mm/s, and $\tau_{coll}/\tau=1.23 \times
10^3$. This means that collisions are rather rare events compared to
the stopping time, and rotations induced by collisions are therefore
damped very quickly. This remains valid as the dust to gas ratio at
the midplane of the disk is increased, until the dust-to-gas mass
ratio approaches unity, at which point dust motion begins to dominate
the disk dynamics. It is therefore clear that collisions are not an
obstacle to photophoretic motion.

Irregularly shaped dust particles can also be subject to torques
induced by gas drag, by interaction with incident radiation, 
such as radiation pressure, or by the effect of photophoresis
itself. Rotations are induced whenever one or more of these external
forces is not in line with the center of mass of a dust grain. All
these effects, however, cause rotations around axes aligned radially
in a protoplanetary disk, as we will outline here, and can therefore be shown to not affect the
photophoretic drift. Weidenschilling (\cite{weidenschilling1977,
weidenschilling1997}) showed that, in the mid-plane of a gaseous
protoplanetary disk, dust particles with gas grain coupling times
smaller than their orbital period do {\it not} have considerable
velocities relative to the gas in the direction of their orbital
motion, but rather corotate with the gas on a sub-Keplerian orbit.  A
significant transverse drift motion relative to the gas only occurs
for bodies much larger than those we consider. In the radial
direction, however, small particles can attain considerable velocities
relative to the gas, depending on the interplay of the radial forces
at work (described above). Irregularly shaped dust particles that move
through a gaseous environment typically show continuous rotation but
only around an axis parallel to the direction of their motion relative
to the gas. They are aligned along the direction of their motion
relative to the gas.  This was observed in sedimentation experiments
with dust aggregates by Wurm \& Blum (\cite{wurm2000}) for large
Knudsen numbers $Kn = 2 .. 20$.  This can be understood by the simple
geometric reasoning that, at large Knudsen numbers, free molecules
impinging upon a particle surface from one side will exert a
torque only until the center of particle cross section and the center
of mass are aligned along the line of gas-particle motion. It should be
pointed out that a high number of gas molecules is needed to set a particle
in rotation. So any effects due to fluctuations of the molecule impacts are
statistically washed out. Brenner
(\cite{brenner1963}) also showed theoretically that in the Stokes
regime, at small Knudsen numbers, the rotation axis of an arbitrary
body moving through a gas will lie parallel to the direction of its
translational motion due to the torques exerted upon it by the gas.

The effect of rotation alignment due to incident radiation was
investigated experimentally by Abbas et al. (\cite{abbas2004}). They
illuminated nonspherical SiC particles of a few microns diameter which
were levitated in an electrodynamic trap with a laser beam and,
thereby, forced them into a stable rotational state around an axis
parallel to the incident laser beam. In a protoplanetary disk, this
would correspond to the radial alignment of rotation axes of those
dust grains which are small enough to be efficiently influenced by
radiative forces.

Thus, we can conclude that, in general, the rotations of the dust
particles (smaller than 10 cm) in a protoplanetary disk are either too
slow, or have rotation axes aligned in such a way that particle
rotation is not an obstacle to photophoresis. However, even
unfavourable rotation (such as that of larger bodies or of
smaller particles in the short time after a collision) will not
completely undo photophoresis. We plan experimental studies to
investigate the effect of particle rotation on the efficiency of
photophoresis in the near future.

\section{Caveats and future work}

There are some aspects of our model which are simplified and need to
be treated in more detail in the future. The most important is the
size distribution of particles in the disk, with particular emphasis
on the smallest particle size that is abundant enough to produce a
high opacity. While we believe that our choice of Eq. (\ref{size}) for
the minimum particle size evolution is reasonable, other evolutions
can drastically change the motion of the dusty edge, delaying, or
speeding up, every part of the evolution. The calculations assume that
the dust can be treated as a trace particle with respect to the
gas. As a strong density enhancement results at the edge, this
assumption might no longer be valid. If the dust density becomes
comparable to, or larger than, the gas density, the drift dynamics
will change. The dust will move on paths which ever more closely
resemble Keplerian orbits, and will carry the gas along with it. The
inward drift will get smaller. This is not important for the
consideration of the initial inner clearing, but the equilibrium
points will shift outwards and the edge might therefore stop its
outward motion much further out. However, as a strong particle
concentration influences the formation of planetesimals and planets,
this is an interesting avenue for further work. Turbulent
diffusion might act to smear out the dust distribution at the
equilibrium positions in an optically thin disk. It has pronounced
influence on the formation and size of the central clearing of the
disk. A more rigorous treatment of turbulent diffusion is needed.
Also, we describe an optically thin disk with an optically thick
model. For the calculation of the behaviour of the outer part of the
disk, and the drift of the edge itself, this might be justified, but
the disk will change after a gap is cleared. This will not, however,
change the principal mechanisms of photophoretic drift. Certainly, the
clearing of gaps by planets is entirely possible, and might influence
the progression of the initial clearing phase. It should also be noted
that our model does not take into account the attenuation of the
stellar radiation due to Rayleigh scattering by the gas molecules in
the disk, which might reduce the radial drift velocities and, thus,
change the positions of the inner disk edge and the equilibrium
points. However, the inner holes of transition disks are currently
modelled best as being transparent (Calvet et al. \cite{calvet2002};
Forrest et al. \cite{forrest2004}; D'Alessio et
al. \cite{dalessio2005}; Bergin et al. \cite{bergin2004}; Calvet et
al. \cite{calvet2005}). For such systems extinction will be only of
minor concern for photophoretic effects at the edge.

A lot of unknown properties of the particles with respect to
photophoresis are currently hidden in the factor $J_1$ in
Eq. (\ref{Fph}). This concerns both the optical properties of grains
and heat transfer. It should also be mentioned that particles can
experience negative photophoresis, which means that particles can be
attracted by light. However, this is restricted to the smallest
particles, roughly below $\rm 10 \mu m$ in size (Mackowski
\cite{mackowski1989}; Rohatschek \cite{rohatschek1985}). The fraction
of the disk made up of such small particles (which exhibit either
negative photophoresis or no photophoresis) might delay the clearing
of the disk until larger particles have grown. However, the drift of
the edge is slow for particles smaller than $\rm 10 \mu m$, and this
will also not change the principle of photophoretic clearing. A better
understanding of the thermal conductivity of large dust aggregates is
also needed.  With respect to the possible applications mentioned in
Sect. 6 it should be pointed out that we do not know the precise point
in time at which the photophoretic sweep of dust described in this
paper starts. The aspect of the chronological relation of the
photophoretic dust migration to other events and epochs in the
evolution of the Solar nebula should be adressed in future studies.
Despite these apparent limitations, and the number of parameters which
are still poorly defined or unknown, we regard photophoretic clearing
as a mechanism that is very likely to be important in the formation of
both our Solar system, and those around other stars.

\begin{acknowledgements}
O. Krauss and G. Wurm are funded by the Deutsche
Forschungsgemeinschaft, and Jonti Horner acknowledges the support of
the Swiss National Science Foundation.

\end{acknowledgements}

\end{document}